\documentclass[aps,twocolumn,pra]{revtex4}

\usepackage{epsf}
\usepackage[dvips]{epsfig}

\date{\today}
\begin{document}

\title{ New possible source of huge neutrino bursts}

\author{S. E. Juralevich and V.V.Tikhomirov \footnote {tikh@inp.minsk.by}}
\address
{Institute for Nuclear Problems, Belarussian State University,
Bobruiskaya 11, 220050 Minsk, Belarus}

 \draft
\begin{abstract}

It is shown that primordial black holes (PBHs) of masses $M\geq
5\times 10^{14}g$ are able to absorb white dwarfs (WD) completely
for the time of their existence, giving rise to formation of
black holes of WD mases. The WD absorption is accompanied by up to
$10^{52}erg$ neutrino bursts which could both considerably
contribute to the cosmic neutrino flux and open up a new way of
PBHs detection, especially effective in placing new constraints on
abundance of nonevaporating PBHs with masses $M\gg 10^{15}g$.
\end{abstract}

 \maketitle

PBHs could have formed in the early universe \cite {zel,haw1}
from initial inhomogeneities, topological defects (strings,
bubbles) or at cosmological phase transitions. PBHs provide a
unique probe of the Big Bang and even their non-existence gives
vital cosmological information \cite {car1,khl,nov}.

The Hawking radiation \cite {haw2}  allows to set up severe
constraints on the PBH abundance. In particular, the radiation of
all the evaporated PBHs could have given its contribution to the
diffuse cosmic flux \cite {car2,hal,bug}. Very stringent bound
has been obtained from the absence of a detectable diffuse flux
of 100 MeV photons \cite {car2}. It was claimed, however, that
these bounds can be improved using available data on diffuse
neutrino flux \cite {hal,bug}.

Another way of PBHs search consists in attempts to detect gamma
\cite {car2} or neutrino \cite {hal} bursts accompanying the
final stage of PBH evaporation. The PBHs which complete their
evaporation at present have had the initial mass $M_0$ equal to
the critical one (Hawking mass) $M_{\ast}\simeq 5 \times
10^{14}g$. For all the time of their evaporation such PBHs could
emit about $M_{\ast}c^2\simeq 5\times 10^{35}erg$, where $c$ is
the speed of light, what is much less, for example, than the
energy of supernova explosion reaching $3\times 10^{53}erg$. The
comparatively low energy release undoubtedly constrains the search
possibilities of the most intensively evaporating PBHs and makes
utterly ineffective that of weakly evaporating PBHs with $M\simeq
M_0 \gg M_{\ast}$, the constraints on which are more than ten
orders lower than that on PBHs with $M_0 \simeq M_{\ast}$ \cite
{car1,khl,nov}.

We would like to suggest here an alternative method of PBHs
search, free of this deficiency. Namely, we will show that being
trapped inside a white dwarf (WD) a PBH of mass $M_0 \geq
M_{\ast}$ can completely absorb it initiating a huge neutrino
burst with the energy reaching $10^{52}erg$.

First, we will consider the simplest case of nonrotating cool WD
and apply the general relativistic theory of spherical,
steady-state, adiabatic accretion of continuous Pascal liquid
\cite {sha,mic} to show that the time of a complete WD absorption
by a Schwarzschild PBH can be less than the age of the universe.
A special consideration has shown that the heating caused by PBH
Hawking radiation, WD matter compression and neutronization does
not violate the high degeneracy of WD matter. The mass-energy
density $E$ and pressure $P$ of the degenerate WD matter  are
given by the equations \cite {sha,mic,lan}
\begin{eqnarray}%1
 \nonumber
E(n)=\mu c^2n+\frac{m^4_ec^5}{8{\pi}^2 {\hbar}^3}\biggl[x(2x^2+1)\sqrt{1+x^2} \\
-\mbox{Arsh}x\biggr]-\frac{3}{4}{\alpha}^{\prime}m_e^2c^3x,
\end {eqnarray}
\begin {equation}%2
P(n)=n\frac{dE}{dn}-E(n) ,
\end{equation}
where $n$ is the electron number density,
\begin {equation}%3
x=x(n)=\frac{p_F(n)}{m_ec}
\end{equation}
is the electron Fermi momentum $p_F(n)=(3{\pi}^2n)^{1/3}\hbar$
expressed in $m_ec$ units, $m_e$ and $\hbar$ are the electron
mass and the Planck constant, respectively. $\mu$ is the WD
matter mass which falls at one electron. We will consider the
most widespread case of carbon WDs for which $\mu\simeq 2m_n$,
where $m_n$ is the nucleon mass. The term of Eq.(1) containing
the small parameter
\begin {equation}%4
{\alpha}^{\prime}=\frac{4}{5}\biggl( \frac{3}{2\pi}\biggr)^{1/3}\alpha Z^{2/3}\ll 1,
\end{equation}
where $\alpha\simeq 1/137$ and $Z$ is the atomic number,
describes the Coulomb correction \cite {sha} which is principally
important in the case of low WD density.

The spherical, steady-state accretion flow of degenerate Fermi
gas is described by the radial component of 4-velocity $u$ and
the electron number density $n$. Both of them depend only on the
radial Schwarzschild coordinate measured from the PBH center and
obey \cite {sha,mic} the electron conservation equation
\begin {equation}%5
 \dot M=4\pi\mu n(r)cu(r)r^2=\mbox{const}
\end {equation}
which gives the rest mass accretion rate as well as the relativistic Bernoulli equation
\begin{eqnarray}%6
 \nonumber
\biggl(\frac{E(n)+P(n)}{\mu c^2n}\biggr)^2\biggl(1+u^2-\frac{r_g}{r}\biggr) \\
=\biggl(\frac{E(n_0)+P(n_0)}{\mu c^2n_0}\biggr)^2=\mbox{const}.
\end {eqnarray}
containing the gravitational radius $r_g=2GM/c^2\simeq
1.5(M/10^{15}g)10^{-15}m$ where $G$ is the Newton's constant. The
steady-state accretion assumes that $n\to n_0=\mbox{const}$ at
$r\to\infty$. The constant signs in Eqs. (5) and (6) reflect the
independence on $r$.

The basic point of the theory \cite {sha,mic} consists in
existence of a special (sonic) point $r=r_s$ at which the
relations
\begin{equation}%7
u_s^2=u^2(r_s)=\frac{a_s^2}{1+3a_s^2}=\frac{r_g}{4r_s}
\end{equation}
hold where  $a_s=a(r_s)$ is the sound speed $a=(dP/dE)^{1/2}$
expressed in $c$ units and taken at $r=r_s$. Using the Emden
solution \cite {sha,lan} for the WD density distribution one can
make sure that, if a PBH with mass $M$ is situated at the WD
center, the uniformity condition in the essential accretion
region $r\sim r_s$ is fulfilled at $M\leq 10^{-3}M_{WD}$ where
$M_{WD}$ is the WD mass. The electron number density $n_0$ in Eq.
(6) can be considered as that at the WD center in this case.

Substituting Eqs. (1), (2) and (7) into Eq. (6) one obtains in
the first order with respect to $m_e/\mu\simeq 2.74\times
10^{-4}$ the equation
\begin {equation}%8
\frac{2+x_s^2}{\sqrt{1+x_s^2}}-{\alpha}^{\prime}x_s=2\biggl(\sqrt{1+x_0^2}-{\alpha}^{\prime}x_0\biggr)
\end{equation}
connecting $x_s=x(n(r_s))$ with $x_0=x(n_0)$. Eq. (8)
demonstrates that the Coulomb correction is indeed essential at
the density $\rho =E/c^2\leq 100 g/cm^3$ giving rise to the
existence of two solutions of Eq. (8) at $x_0<2{\alpha}^{\prime}$
(${\alpha}^{\prime}\simeq 0.015$ at $Z=6$). However here we will
consider the usual case of much denser WDs in which $x_0\gg
2{\alpha}^{\prime}$ and Eq. (8) has a single solution taking the
simple form
\begin {equation}%9
x_s\simeq 2x_0+\frac{1}{4(1-{\alpha}^{\prime})x_0}+O\biggl(\frac{1}{{x_0}^2}\biggr)
\end{equation}
or $\rho (r_s)\simeq 8 {\rho}_0$ in the ultrarelativistic limit
of $x_0\gg 1$, $\rho _0\gg 2\times 10^6g/cm^3$. Eqs. (5) and (6)
also allow one to calculate the parameter (3) maximum value
\begin{equation}%10
x_{max}=x(r_g)\simeq\frac{(1+x_s^2)^{1/4}}{2^{4/3}}\biggl(\frac{3\mu}{m_e}\biggr)^{1/2}\simeq
42(1+x_s^2)^{1/4}
\end {equation}
at $r=r_g$. The corresponding Fermi energy
${\varepsilon}_F(r_g)\simeq m_ecx(r_g)$ reaches $20-100 MeV$.

Substituting $x_s$ into Eq. (7) one finds the values $n(r_s)$,
$r_s$ and $u(r_s)$ allowing to calculate the right hand side of
Eq. (5) and to obtain the law
\begin{equation}%11
M(t)=\frac{M_0}{1-t/T_{abs}}
\end {equation}
of the PBH accretion mass growth starting from its initial value
$M_0=M(0)$ and characterized by the time
\begin{eqnarray}%12
 \nonumber
T_{abs}=\frac{\pi {\hbar}^3}{\sqrt{3}G^2{\mu}^{5/2}m_e^{3/2}M_0}\biggl(\frac{1}{\sqrt{1+x_s^2}}-\frac{{\alpha}^{\prime}}{x_s}\biggr)^{3/2} \\
=\frac{27}{(M_0/10^{15}g)}\biggr(\frac{1}{\sqrt{1+x_s^2}}-\frac{{\alpha}^{\prime}}{x_s}\biggl)^{3/2}
Gyr.
\end {eqnarray}
which can be considered as that of WD absorption. It should be
emphasized that though Eq. (11) loses its applicability at
$M(t)\geq 10^{-3}M_{WD}$, the accuracy of the estimate (12) is
quite high since Eq. (11) does not describe only a very small part
$\Delta t/T_{abs}=M_0/(10^{-3}M_{WD})$ ($10^{-16}$-th part at
$M_0\sim M_{\ast}$) of the time (12). Thus, the approach \cite
{sha,mic} has led us to a very precise estimate of the time of WD
absorption by a PBH.

Using the relation (8) of $x_s$ with $x_0=x(n_0)$ one can
calculate the time (12) dependence on the WD central density
$\rho _0=E(n_0)/c^2$. Fig. 1 demonstrates that a PBH with initial
mass $M_0=10^{15}g$ can absorb relativistic ($\rho _0\geq 10^6
g/cm^3$) WDs for several $Gyrs$. Since $T_{abs}\propto M_0^{-1}$,
PBHs with initial masses $M_0\geq 10^{17}g$ will absorb WDs of any
density for the time of their existence. The absorption process
could give rise to formation of black holes of WDs masses which
can not form in the star collapse process. Since WDs are quite
abundant their absorption by PBHs could, in principle, be more
productive source of star-mass black holes than the star collapse.

\begin{figure}[!ht]
\centering \psfull
    \epsfig{file=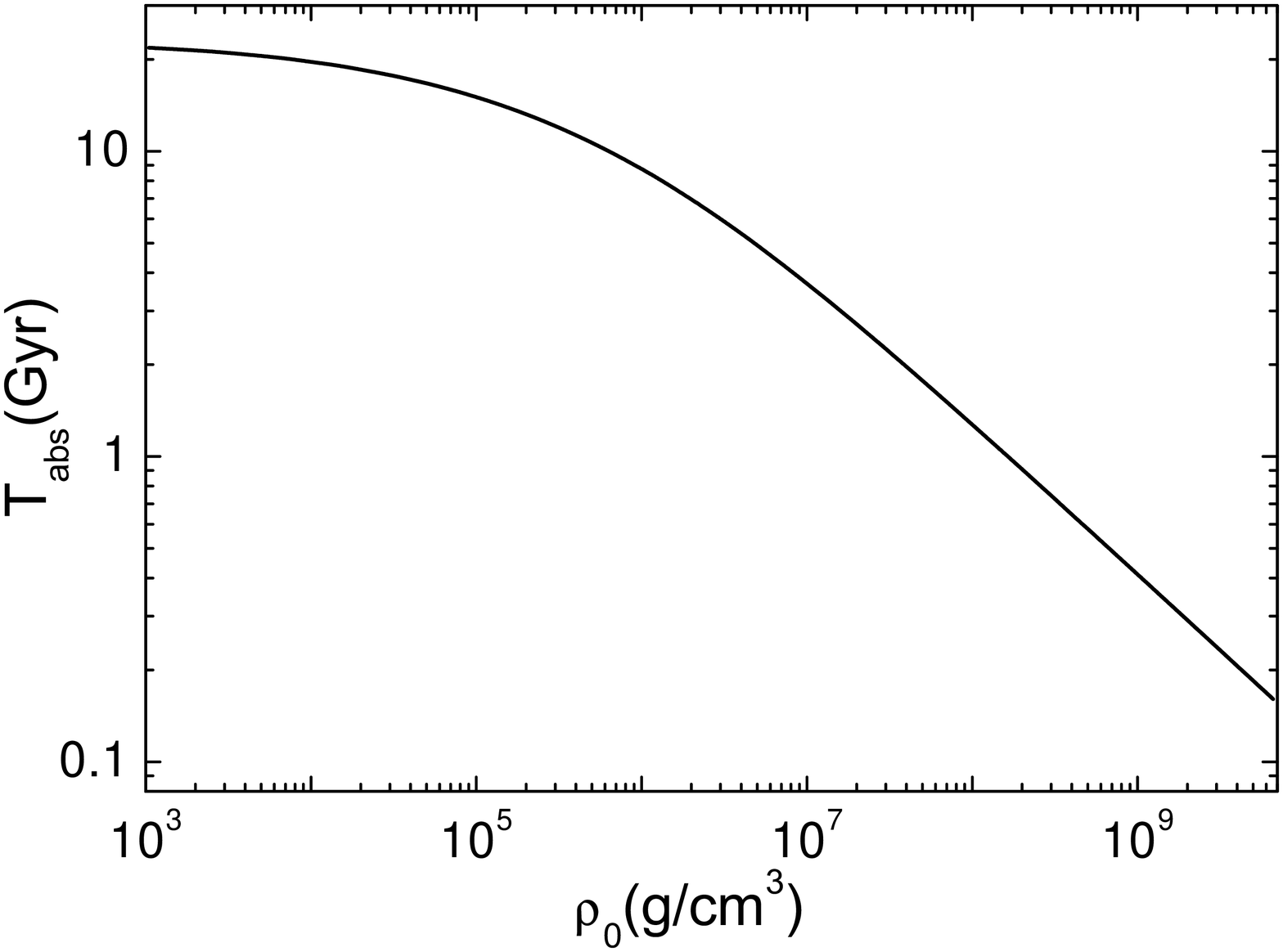,width=8cm}
    \caption{Time of a complete WD absorption by a PBH with mass $M_0=10^{15}g$ as a function
     of WD central density.}
\end{figure}

A WD absorption by a PBH would be observed as its sudden
disappearance accompanied by a moderate light ($x$-ray) flash.
Such a disappearance, however, is difficult to observe at
distances exceeding $100ps$. We will show now that a WD
absorption by a PBH is accompanied by a huge neutrino burst
detectable at much longer distances.

Neutrinos will be produced in neutronization reactions
($^{12}C+e^-$$\to$$ ^{12}B+\nu$, $^{12}B+e^-$$\to$$ ^{12}Be+\nu$
in carbon WDs \cite {sha}) in the region with electron Fermi
energy ${\varepsilon}_F(r)=m_ec^2\sqrt{1+x^2(r)}$ exceeding the
neutronization threshold energy $\Delta _{\mbox{\tiny Z}}$
($\Delta _{\mbox{\tiny C}}=13.88 MeV$, $\Delta _{\mbox{\tiny
B}}=12.17 MeV$). Since $\varepsilon _F(r_g)$ exceeds $\Delta
_{\mbox{\tiny C}}$ considerably at any $x_0$ and $\rho _0$ (see
Eqs. (8) and (10)) a PBH will give rise to the neutronization
process in any carbon WD.

Both the local neutronization probability of a nucleus with atomic number $Z$ \cite {sha}
\begin{equation}%13
\Gamma _{\mbox{\tiny Z}}=\Gamma _{\mbox{\tiny Z}}(\varepsilon _F(r))=\frac{ln2}{T_{\mbox{\tiny {Z-1}}}} \frac{f(\Delta _{\mbox{\tiny Z}}, \varepsilon _F)}{f(m_ec^2, \Delta _{\mbox{\tiny Z}})},
\end {equation}
where
\begin{eqnarray}%14
 \nonumber
f(\varepsilon _1, \varepsilon _2)=\int_{\varepsilon _1}^{\varepsilon _2} f^\prime d\varepsilon, \\
f^\prime =\varepsilon (\varepsilon ^2-m_e^2c^4)^{1/2}(\varepsilon
-\Delta _{\mbox{\tiny Z}})^2 ,
\end {eqnarray}
and the accompanying neutrino emission intensity
\begin{eqnarray}%15
 \nonumber
I^{\mbox{\tiny Z}}_{\nu}=I^{\mbox{\tiny Z}}_{\nu}(\varepsilon _F(r))=\frac{ln2}{T_{\mbox{\tiny {Z-1}}}f(m_ec^2,\Delta _{\mbox{\tiny Z}})} \\
 \times \int _{\Delta _{\mbox{\tiny Z}}}^{\varepsilon _F}(\varepsilon -\Delta _{\mbox{\tiny Z}})f^\prime d\varepsilon.
\end {eqnarray}
are determined by the local electron number density $n(r)$, the
threshold energy $\Delta _{\mbox{\tiny Z}}$ and the half-life
$T_{\mbox{\tiny {Z-1}}}$ of the daughter nucleus ($T_{\mbox{\tiny
B}}=20.20 ms$, $T_{\mbox{\tiny {Be}}}=21.31 ms$).

Neutronization begins at the "neutronization radius" $r_n > r_g$
at which $\varepsilon _F(n(r_n))=\Delta _{\mbox{\tiny C}}$ and
$\rho _n=E(n(r_n))/c^2$ $=3.90\times 10^{10}g/cm^3$ \cite {sha}.
According to Eq. (9) one has $r_n = r_s$ at the central WD
density $\rho_0 \simeq \rho _n/8 \simeq 4.90\times 10^9 g/cm^3$.
Since most of WDs have much  smaller ones let us assume that $r_n
\ll r_s$. The WD matter motion in the neutronization region $r_g
< r < r_n$  is determined in this case rather by gravity than by
pressure and is well described  by the free fall equation
\begin{equation}%16
u(r)=\sqrt{r_g/r}
\end {equation}
written in Lagrangian coordinates. Eq. (16) allows both to
introduce a proper time $\tau (r)=\int _r^{r_n}dr/[cu(r)]$, $\tau
(r_n)=0$, of elementary volume motion from $r_n$ to $r$ and to
write the electron number conservation equation
$r^2nu=\mbox{const}$, which holds in neglect of electron
absorption, in the form
\begin{equation}%17
x(r)=x_{max}\sqrt{r_g/r}.
\end {equation}

Eqs. (8) and (10) obtained in the model of steady-state accretion
\cite {sha,mic} fail to describe the accretion rate saturation
accompanied by the most intense neutrino emission. The last comes
from the region $r\sim r_{max}$ giving the main contribution to
the integral determining the WD mass. According to the Emden
solution for ultrarelativistic Fermi gas \cite {sha,lan}
$r_{max}$ equals $0.237 R_{WD}$, where $R_{WD}$ is the WD radius.
The WD matter with the initial density $\rho(r _{max})=0.312\rho
_0$  experiences the attraction of the mass $0.37 M_{WD}$. Though
this mass is contracted in the process of absorption by a PBH it
determines the constant value of $r_g$ in Eqs. (16) and (17)
describing the final stage of accretion accompanied by
neutronization. The nonuniformity of the region $r\sim r_{max}$
giving the main contribution to the neutrino emission will be
taken into consideration by using $\rho(r_{max})$ instead of
$\rho_0$ in Eqs. (8) and (10) determining the $x_{max}$ value.

\begin{figure}[!ht]
\centering \psfull
    \epsfig{file=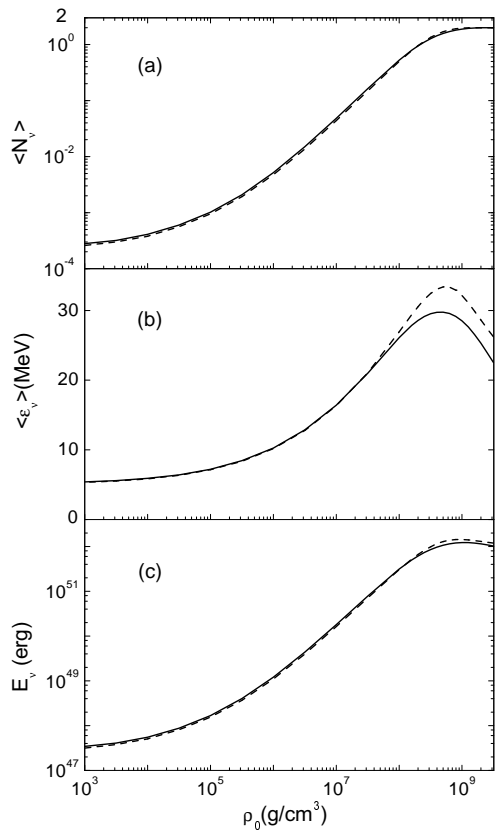,width=8cm}
    \caption{Characteristics of neutrino emission process accompanying a carbon WD absorption by
 a PBH: (a) average  neutrino number which falls at one nucleus, (b) average neutrino energy
  and (c) neutrino burst energy as a function of central WD density as predicted by the model
   described in the text (dashed line) and by a more sophisticated approach (full lines).}
\end{figure}

Let us introduce the $C$, $B$ and $Be$ nuclei fractions
normalized by the equality $c_{\mbox{\tiny C}}+c_{\mbox{\tiny
B}}+c_{\mbox{\tiny {Be}}}=1$ and satisfying the initial
conditions $c_{\mbox{\tiny C}}(0)=1$, $c_{\mbox{\tiny
B}}(0)=c_{\mbox{\tiny {Be}}}(0)=0$ at the moment $\tau (r_n)=0$
of passage through the neutronization radius. The first two
fractions obey the equations
\begin{eqnarray}%18
 \nonumber
dc_{\mbox{\tiny C}}/d\tau =-\Gamma _{\mbox{\tiny C}}c_{\mbox{\tiny C}} \\
dc_{\mbox{\tiny B}}/d\tau =\Gamma _{\mbox{\tiny C}} c_{\mbox{\tiny C}}-\Gamma _{\mbox{\tiny B}}c_{\mbox{\tiny B}}.
\end {eqnarray}
Eqs. (16) and (17) allow to pass to the variable $r$ in Eqs. (13)
and (18) and easily integrate the last yielding the analytical
expressions for $c_{\mbox{\tiny C}}(r)$ and $c_{\mbox{\tiny
B}}(r)$, as well as for both the average neutrino number
\begin{equation}%19
<N_{\nu}>=2-2c_{\mbox{\tiny C}}(r_g)-c_{\mbox{\tiny B}}(r_g)
\end {equation}
which falls at one carbon nucleus and the average neutrino energy

\begin{eqnarray}%20
 \nonumber
<\varepsilon _{\nu}>=\frac{1}{<N_{\nu}>}\int _{r_g}^{r_n}\biggl[ I^{\mbox{\tiny C}}_{\nu}(r)c_{\mbox{\tiny C}}(r) \\
+I^{\mbox{\tiny B}}_{\nu}(r)c_{\mbox{\tiny
B}}(r)\biggr]\frac{dr}{cu(r)}.
\end {eqnarray}
The $<N_{\nu}>$ and $<\varepsilon _{\nu}>$ dependencies on the
central  WD density $\rho _0$ are given, respectively, on Figs.
2(a) and (b) along with the same dependencies evaluated using a
more sophisticated model taking into account the electron
absorption, the pressure influence on accreting matter motion and
the nonequilibrium heating of the last which accompanies the
neutronization process \cite {bys}. The negligible difference in
predictions of these two models justifies the undertaken
consideration of the simpler one.

Fig. 2(a) demonstrates the possibility of complete carbon and
subsequent boron neutronization. The average and maximum neutrino
energies can reach 30 (see Fig. 2(b)) and $100 MeV$ (see Eqs. (9)
and (10)), respectively. The dependence of the energy
\begin{equation}%21
E_{\nu}=\frac{M_{WD}}{Z\mu}<N_{\nu}><\varepsilon _{\nu}>
\end {equation}
of the neutrino burst accompanying the total WD absorption on the
central WD density is given on Fig. 2(c) demonstrating that
$E_{\nu}$ can exceed $10^{52} erg$. The neutrino burst duration
will not considerably exceed the free fall time, since neutrino
will travel through the WD matter nearly freely. The huge
neutrino flux, nevertheless, will heat the WD matter up to $T\geq
10^7K$ and, possibly, initiate the nuclear burning \cite {ger} in
the most dense WDs. These processes will be considered elsewhere.

Thus, we have demonstrated that the energy of a purely neutrino
burst accompanying a WD absorption by a PBH is only a few orders
less than that of a neutrino-antineutrino burst accompanying a
supernova explosion \cite {kla}. If such a burst occurs at the
typical distance of $10kps$ it will cause up to 50 events of $\nu
e$ scattering in the Super-Kamiokande detector and, thus, can be
detected with confidence. In fact, it looks feasible to design a
detector able to detect most of the predicted neutrino bursts in
our Galaxy \cite {hal,kla}. A possible frequency of these bursts
depends on such uncertain factors as PBH abundance and mass
distribution as well as on the probability of PBHs presence in
WDs or capture by them. Taking into consideration the high WDs
abundance in the Galaxy and assuming more or less uniform
distribution of the moments of WDs complete absorption in time,
optimistically one can hypothesize that the predicted neutrino
bursts can be detected nearly every year.

According to Ref. \cite {bug} PBHs with masses $M_0\ll M_{\ast}$
contribute mostly to the diffuse neutrino flux, the study of
which already allows to set up restrictions on the power of the
"blue" spectrum of initial density fluctuations. The possibility
of intensive neutrino emission by WDs being absorbed by PBHs
radically change this picture. First, since the energy (21) 11-16
orders exceeds the rest energy $M_{\ast}c^2$, even a small number
of PBHs absorbing most weakly rotating WDs can improve the bounds
on PBHs absence by many orders. Second, since the heavier PBHs
absorb WDs more readily, the diffuse neutrino flux, in fact,
allows to set up the most stringent bounds on the
$nonevaporating$ PBHs with masses $M_0\gg M_{\ast}$.

Thus, we have demonstrated that PBHs can absorb WDs for the time
of their existence giving rise to formation of black holes of
WD's masses. The WDs absorption is accompanied by the neutrino
bursts which can be detected directly and could have given a
considerable contribution to the diffuse neutrino flux. Most
considerably these neutrino bursts widen the possibilities of
search of nonevaporating PBHs with masses $M_0\gg 10^{15}g$.

The authors would like to thank Professor M. Ju. Khlopov and the
members of his seminar for valuable discussion and Professor V.
G. Baryshevsky for support of our work.

\end{document}